# Relativistic interaction of long-wavelength ultrashort laser pulses with nanowires

Zhanna Samsonova[1,2], Sebastian Höfer[1], Vural Kaymak[3], Skirmantas Ališauskas[4], Valentina Shumakova[4], Audrius Pugžlys[4], Andrius Baltuška[4], Thomas Siefke[5], Stefanie Kroker[6,7], Alexander Pukhov[3], Olga Rosmej[8,9], Ingo Uschmann[1,2], Christian Spielmann[1,2], and Daniil Kartashov[1]

1. Institute of Optics and Quantum Electronics, Abbe Center of Photonics, Friedrich-Schiller-University Jena, Max-Wien-Platz 1, 07743 Jena, Germany
2. Helmholtz Institute Jena, Fröbelstieg 3, 07743 Jena, Germany
3. Institute for Theoretical Physics, Heinrich-Heine-University Düsseldorf, Universitätsstraße 1, 40225 Düsseldorf, Germany
4. Photonics Institute, Vienna University of Technology, Gußhausstraße 27-29, 1040 Vienna, Austria
5. Institute of Applied Physics, Abbe Center of Photonics, Friedrich-Schiller-University Jena, Albert Einstein Straße 15, 07745 Jena, Germany
6. Laboratory for Emerging Nanometrology, Technical University Braunschweig, Pockelsstraße 14, 38106 Braunschweig, Germany
7. Physikalisch-Technische Bundesanstalt, Bundesallee 100, 38116 Braunschweig, Germany
8. GSI Helmholtz Centre for Heavy Ion Research, Planckstraße 1, 64220 Darmstadt, Germany
9. Goethe-University, Institute of Applied Physics, Max-von-Laue-Straße 1, 60438 Frankfurt am Main, Germany

**Abstract**

We report on experimental results in a new regime of a relativistic light-matter interaction employing mid-infrared (3.9 μm wavelength) high intensity femtosecond laser pulses. In the laser generated plasma, the electrons reach relativistic energies already at rather low intensities due to the fortunate $\lambda^2$-scaling of the kinetic energy with the laser wavelength. The lower intensity suppresses optical field ionization and creation of the pre-plasma at the rising edge of the laser pulse efficiently, enabling an enhanced efficient vacuum heating of the plasma. The lower critical plasma density for long-wavelength radiation can be surmounted by using nanowires instead of flat targets. In our experiments ≈80% of the incident laser energy has been absorbed resulting in a long living, keV-temperature, high-charge state plasma with a density of more than three orders of magnitude above the critical value. Our results pave the way to laser-driven experiments on laboratory astrophysics and nuclear physics at a high repetition rate.

## I. Introduction

Solid density, (multi-)keV temperature plasmas open new perspectives for realizing table-top, high brilliance X-ray sources [1-3], laser-induced nuclear physics [4, 5] and experiments on laboratory astrophysics [6, 7]. Such plasmas can be generated when relativistically intense, high temporal contrast femtosecond laser pulses interact with solids. In the relativistic regime of laser-matter interaction the velocity of the laser field driven free electrons approaches the speed of light. Quantitatively, it is reached, if the normalized vector potential $a_0 \geq 1$, where $a_0 = 0.85 \times 10^{-9}\sqrt{I}\lambda$, with $I$ is the laser pulse peak intensity in W/cm² and $\lambda$ is the laser wavelength in μm. A common way to enter the relativistic regime is to increase the laser intensity. For a given laser pulse energy, higher intensities can be reached easier with short wavelength laser sources by entering the so-called $\lambda^3$ regime, defined by focusing of the shortest possible pulses (given by the length of single-cycle) to the diffraction limited spot given by the wavelength [8]. However, such single-cycle ultra-intense pulses will only interact with an extremely small volume of the plasma and will be not able to heat it to high temperatures. It is also worth mentioning that pulse compression down to a single cycle becomes extremely challenging with a shortening of the laser wavelength.

Alternatively, $a_0$ can be more efficiently increased using long wavelength laser sources. In this case, highly relativistic interaction can be realized with moderate laser pulse energies paving the way to high repetition rate experiments in a relatively large volume. This new regime of relativistic, ultra-short pulse laser-solid interaction was inaccessible so far, because $CO_2$ lasers, the only high energy, mid-infrared (mid-IR) laser sources available, emit only pulses with a duration >1 ps [9, 10]. For such long durations the hydrodynamic expansion during the pulse prevents plasma densities above the critical value. Due to the recent progress in the development of high energy optical parametric chirped-pulse amplification (OPCPA) femtosecond laser systems, nowadays TW-level peak power femtosecond laser pulses in mid-IR spectral range are available [11, 12].

The requirement of high temporal contrast of a relativistically intense, ultrashort laser pulse is one of the necessary conditions to achieve high density and sharp (on the laser wavelength scale) spatial gradients in plasmas. Sharp gradient of the density enables efficient vacuum heating mechanism of the laser energy absorption [13-17], thus ensuring high temperature of generated dense plasmas. For high power laser systems in near-IR spectral range, commonly used for relativistic laser-solid interactions, the requirement of the high temporal contrast inevitably invokes frequency doubling of the output radiation, reducing the ionization threshold to a few- or even single-photon absorption. Significant reduction of the ionization rate in comparison to near-IR-visible-UV laser sources [18], together with long wavelength in mid-IR laser pulses, substantially relax the problem of pre-plasma formation and achieving sharp spatial gradients of the plasma density. In particular, an advantage of mid-IR femtosecond driver pulses, compared to near-IR pulses, for the efficiency of vacuum heating of solid target under non-relativistic interaction, resulting in a much higher yield of the $K_\alpha$ emission from a Cu tape, was recently demonstrated [19, 20].

Here we report for the very first, to the best of our knowledge, experimental results on a new regime of relativistic laser-solid interaction drawing on the combination of mid-infrared wavelength, high peak power, high temporal contrast femtosecond laser source and advanced nanostructured silicon samples as the target. The measured characteristic line and bremsstrahlung X-ray emission spectra, together with numerical simulations of the laser-plasma interaction and plasma emission, based on the particle-in-cell (PIC) [21] and collisional radiative population kinetics [22] codes, reveal that the nanowire morphology allows overcoming the problem of the unfavorable scaling of the critical electron density with the wavelength $n_{cr}[cm^{-3}] \approx \frac{1.8^{1}}{\lambda^2[\mu m]}$. The generated plasmas have solid density, corresponding to an unprecedented >$10^3 n_{cr}$ of the driving laser pulses.

## II. Experimental setup

The experiments were carried out at the high energy OPCPA laser system delivering 90 fs laser pulses at the 3.9 μm idler wavelength with the energy on the target up to 25 mJ at a 20 Hz repetition rate [11, 12]. The beam was focused by an off-axis parabolic mirror onto a

12 µm (FWHM) focal spot under 45° angle with respect to the target's normal (see Fig. 1a). The pulses were characterized by the SHG FROG measurements and the focal beam size was measured by the knife-edge scan method. The maximum peak intensity at the target was estimated to $10^{17}$ W/cm$^2$ resulting in the maximum value of the relativistic parameter $a_0 \approx 1.1$. All the measurements were conducted in a single shot regime.

In conventional Chirped-Pulse Amplification laser systems based on the laser principle of amplification the temporal contrast at the nanosecond-picosecond time scale is determined by the level of the amplified spontaneous emission and amplification of any parasitic pre-pulses [25]. For OPCPA the temporal structure of the output pulse is mainly defined by the emission of superfluorescence having the duration of the pump laser pulse (80 ps for our system). To get an estimate of the temporal contrast in our experiments, the seed signal beam was blocked to measure the energy of the superfluorescence radiation. The background signal was below the mW level which is the sensitivity threshold of the used detector for 20 Hz repetition rate. This number has to be compared with the amplified idler signal at the 25 mJ output energy resulting in 500 mW measured average power. Thus the energy in the ~80 ps superfluorescence pulse was below 50 µJ and the corresponding peak intensity of ~$10^{11}$ W/cm$^2$ at the 3.9 µm wavelength is too low to generate any significant amount of plasma before the main pulse.

The targets represent a single crystalline Si nanowire arrays (Fig. 1b) on a silicon substrate, transparent in the broad spectral range of 1.2-15 µm. The details of the target fabrication are given in [26]. Briefly, 500 µm-thick Si substrates were coated with a chromium layer, followed by an electron beam sensitive resist. The periodic patterns were created during the electron beam lithography with character projection. After the development of the resist, reactive ion etching was carried out to transfer the structure to the chromium layer and subsequently to the silicon substrate. Finally, residuals of resist and chromium layer were removed using wet chemical etching. The resulting density of the Si NW is about 35 % solid density, which means that a focal spot of 12 µm contains roughly 1600 NWs. Each NW is 6 µm-long and has a diameter of 200 nm. The spacing between the NWs is about 100 nm. The reference target is a 500 µm-thick polished Si wafer. All samples have an area ~1 cm$^2$ allowing several tens of shots per sample with a pretty good reproducibility of the measured X-ray and hard X-ray spectra due to low (below 10%) shot-to-shot fluctuations of the laser energy.

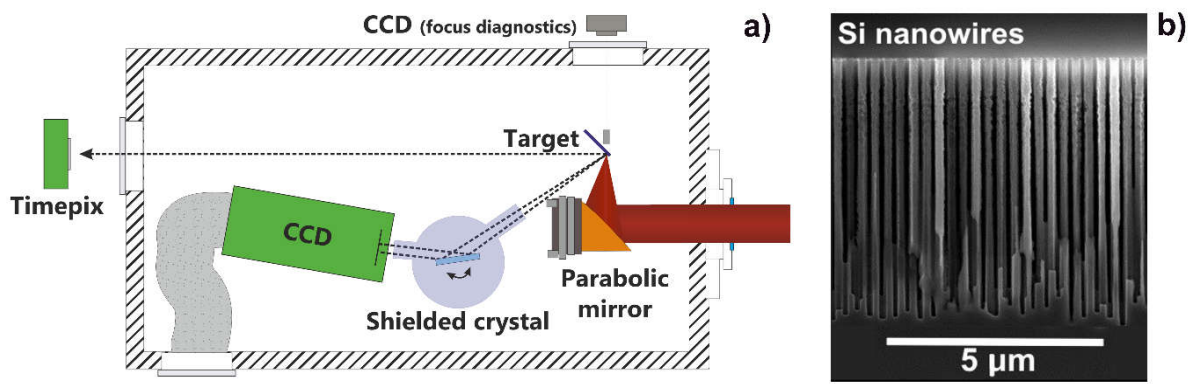

*Fig. 1* (Color online) a) Experimental setup. b) Scanning electron microscope (SEM) image of the Si nanowire target (side view).

The experimental diagnostics include high resolution X-ray crystal spectrometer for the range of 1.7-2.1 keV covering K-shell characteristic emission lines from Si (from $K_\alpha$ to $K_\beta$) for all charge states up to H-like, and a detector for characterizing the hard X-ray - gamma-ray bremsstrahlung emission spectrum. The line emission was registered with a crystal spectrometer based on a flat potassium acid phthalate (KAP) crystal combined with a cooled back-illuminated X-ray CCD camera. The hard X-ray and gamma-ray spectrum in a broad energy range was measured with a Timepix detector, based on a CMOS pixel read-out chip working in a single photon counting mode. A 1 mm-thick CdTe sensor chip enables registration of photons with energies up to 800 keV [27]. The focusing parabola, the targets and the spectrometer were placed in a vacuum chamber pumped below $10^{-4}$ mbar pressure. The Timepix detector was located in ambient air ~5 m away from the target and oriented at 45° to the target normal observing it through a 230 μm-thick Kapton window.

### III. Experimental results

The spectra of the characteristic line emission from the polished and nanowire targets are shown in Fig. 2a. Under conditions of our experiment the efficiency of the $K_\alpha$ emission from weakly ionized Si ("cold" emission), is essentially the same from the both morphologies. This result is strikingly different from the results known in the non-relativistic regime of interaction [28-30] and suggests that in the relativistic limit nanostructure arrays have no advantages in comparison to the flat surface in terms of the efficiency of the characteristic line emission from weakly charged states.

In contrast to the characteristic emission from $Si^+$- $Si^{4+}$ contributing to the $K_\alpha$-line, Fig. 2a demonstrates dramatic increase in the efficiency of the line emission from highly charged ion species for the nanowire arrays in comparison to the flat surface. Emission from Li-like ($Si^{11+}$), He-like ($Si^{12+}$) and H-like ($Si^{13+}$) ions, such as $He_\alpha$, $He_\beta$ and $Ly_\alpha$ is detected from the nanowire arrays only. Note that K-shell emission from $Si^{11+}$ overlaps with $K_\beta$ emission line of $Si^+$ ions, though their maxima are slightly shifted with respect to each other. It is known that the ratio of the intensities of $K_\alpha$ and $K_\beta$ emission lines in Si is about 50:1 [31, 32] which matches well to the measured ratio of 42:1 in the emission spectrum from the polished target (Fig. 2a). In contrast, this ratio for the nanowires is 6:1 proving dominant contribution of the K-shell emission from the Li-like Si charge state in the spectral range 1.83-1.85 keV (the spectral range covering $Si^{11+}$ emission marked in Fig. 2a).

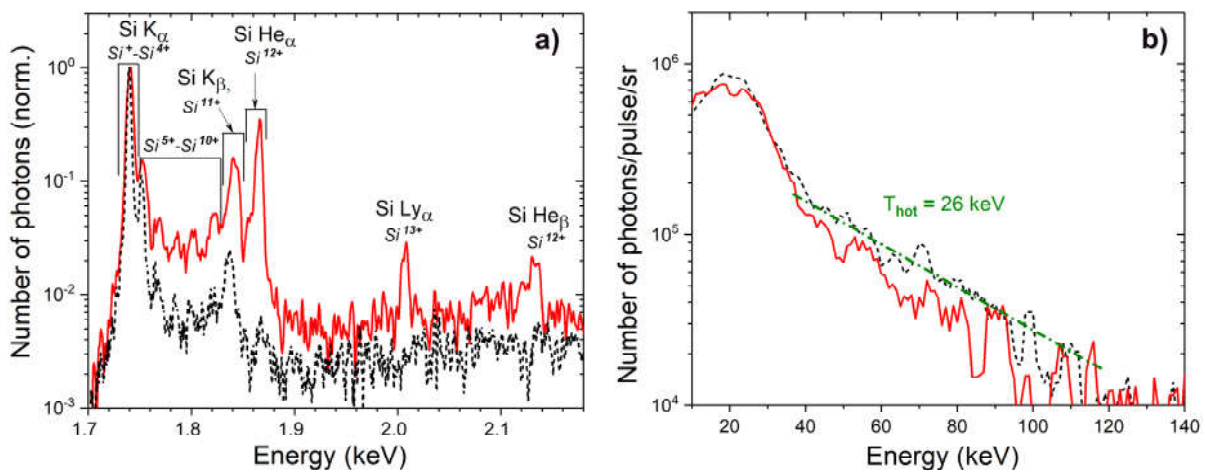

***Fig. 2*** (Color online) a) Spectrum of K-shell line emission from the polished (dotted black line) and the nanowire array (solid red line) Si targets in the logarithmic scale. The identified emission lines contain also groups of K-shell satellites from the intermediate charge states. b) Hard X-ray emission spectrum measured from the polished Si target (black dashed line) and the nanowire array (red solid line). The dashed-dotted green line is the exponential fit providing an estimate of the hot electron temperature $T_{hot}$=26 keV.

Similar to the measurements of "cold" $K_\alpha$ emission, the high energy bremsstrahlung spectrum show almost no difference in the flux up to 140 keV for flat and nanostructured targets (Fig.2b). By applying an exponential fit to the hard X-ray photon distribution [33], for both target morphologies we estimate approximately the same hot electron temperature $T_e \approx 26$ keV for the electron fraction responsible for the emission of the photons in the energy range 40-140 keV. The "hot" electron temperature estimated from the hard X-ray spectrum in Fig. 2b is the temperature of the electron fraction heated during the laser pulse via the Brunel mechanism [13-17]. The general structure of the electron energy distribution function (EEDF) for relativistic laser-plasma interaction is up to now an open question because numerical simulations in the entire energy range, from the relatively cold major part of EEDF to the ultrafast relativistic tail controlling ion acceleration, are extremely challenging. Also it is sensitive to several parameters like the spatial gradient of the plasma density and $a_0$ [13-17]. However, from the general physical point of view, it is reasonable to expect that the EEDF should consist of at least three fractions of electrons. The first one, commonly termed as "hot" electrons in the particle acceleration community, constitutes only a tiny fraction of the total amount of electrons. They are accelerated and pushed out by the ponderomotive force in the laser pulse and leave the target forming the charge separation sheet at the surface [34, 35]. The second fraction, which we call "hot" in this paper, are forming an energetic tail of the EEDF due to the Brunel heating mechanism in the laser pulse. According to the PIC simulations (see below), under the conditions of our experiments this fraction contains less than 1% from the total amount of electrons. Finally, the third fraction, containing the absolute majority of electrons, has the bulk electron temperature which can be estimated from simulations of the measured line emission spectra (see below). Examples of such a three-temperature EEDF are described e.g. in [36, 37].

## IV. Discussions

To get insight into the physics of processes occurring at different time scales under conditions of our experiments, we performed two sets of numerical simulations. First we used particle-in-cell (PIC) simulations using the Virtual Laser Lab code [21] to retrieve the parameters of plasma and their dynamics under relativistic interaction of ultra-high temporal contrast ultrashort laser pulses with Si nanowire arrays. The obtained temperature for hot and bulk electrons and their temporal evolution were used for calculating the line emission spectra using the FLYCHK atomic kinetic code [22]. For FLYCHK we assumed a constant ion density of $5 \times 10^{22}$ cm$^{-3}$ implying as that the NWs stay intact over the simulation time of 1.5 ps, while the electron density evolution was considered by the code.

The spatial distribution of the bulk electron temperature along the wires at different moments of time calculated with the PIC code is shown in Fig. 3a. It reaches the maximum temperature of $\approx 600$ eV within a $\approx 1$ μm thick layer at the tip of the wires during the interaction with the laser pulse. After the laser pulse (t=300 fs after its peak) the entire wire volume is heated almost homogeneously to $\approx 50$ eV.

The spatial distribution of the electron density in the wire volume is shown in Fig.3b. It reaches the maximum value $n_e \sim 6 \times 10^{23}$ cm$^{-3}$, corresponding to fully ionized Si atoms up to $Si^{12+}$ (He-like) charge state, within the hot $\approx 1$ μm-thick upper layer of the wires on the time scale of the laser pulse duration. After the laser pulse (t=300 fs after its peak) the main volume of the wires is ionized with the electron density ~3×10$^{23}$ cm$^{-3}$ and thermalizes to the electron temperature of about 30 eV. Considering the value of the critical density $n_c$≈7×10$^{19}$ cm$^{-3}$ at the 3.9 μm laser wavelength, the simulations suggest that plasma as dense as ≈10$^4$ $n_c$ is generated under the conditions of the experiment.

The ≈1 μm depth of the high density and high temperature plasma region is determined by the penetration depth of the laser pulses into the wire array which, in turn, is limited by the absorption. The simulations predict 76% absorption efficiency of the laser energy which is concentrated mostly within the ≈1 μm layer (for the given 20 mJ laser pulse energy) due to extremely high $n_e/n_c$ ratio.

High temperature and density of plasma enabled by NW morphology plays a key role in reaching high-charge state and efficient emission from He-like and H-like ions. Note that simulations based on Popov-Perelomov-Terentiev (PPT) ionization model [38] show that the optical field ionization for the experimental field parameters is capable to ionize Si atoms up to Si$^{4+}$ charge state only, therefore higher charge states are the result of collisional plasma heating and generated via the electron impact ionization mechanism. The evolution of the charge states, populations of different excited ion states and the corresponding emission spectrum were simulated using the kinetic FLYCHK code with the temporal history of the bulk and hot electron temperatures derived from the PIC simulations. The large volume of the dense plasma predicted by the PIC simulations results in a large wavelength-dependent plasma opacity, causing a strong reabsorption of the generated emission and, therefore, influencing the emission line ratio. To include the effect of opacity, the effective thickness of the emitting plasma layer was used as a fitting parameter to match the simulated spectra to the measured ones. The effective thickness of the emitting plasma layer providing the best match between the simulated and the measured X-ray spectra is found to be ≈0.5 μm. The temporal evolution of the bulk and hot electron temperatures averaged over the volume of the upper 0.5 μm layer of wires is shown in Fig. 4a. The maximum average bulk temperature of ≈0.6 keV is reached by the end of the laser pulse and cools down with ~100 fs time constant. The hot electron temperature reaches its maximum of about 25 keV within the laser pulse duration and vanishes as soon as the laser pulse is off. The calculated maximum is in very good agreement with the hot electron temperature estimated from the measured hard X-ray spectra (see Fig. 2b). The simulated emission spectrum is shown in Fig. 4b and demonstrates an excellent agreement with the most energetic part of the experimental spectrum, consisting of He- and H-like Si emission lines and emitted at the maximum intensity region in the focal distribution. Note that the simulation result is quite sensitive to the changes of the input pa-rameters and varying the electron temperature by more than 10% cannot be simply compensated by a reasonable density adjustment. Therefore, we conclude that the used set of the plasma parameters indeed fits well to the experimental findings. It is worth mentioning also that the temporal evolution of the electron density during the laser pulse, calculated by the FLYCHK code, agrees well with the evolution electron density in the PIC simulations averaged over the 0.5 μm upper. After the laser pulse the electron density predicted by the FLYCHK simulations drops by the factor of 2 within ~200 fs due to efficient recombination processes, which is not included in the PIC simulations.

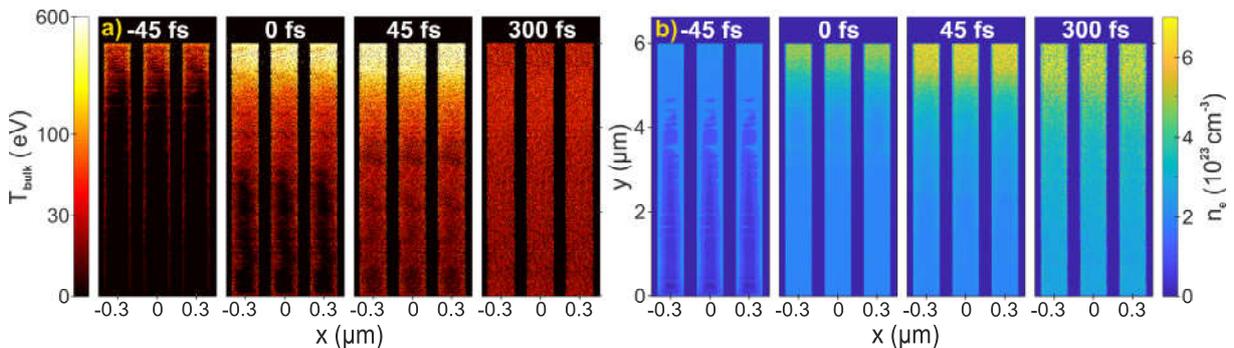

*Fig. 3* (Color on-line) Spatial distribution of a) the electron temperature and b) electron density in Si nanowires retrieved from the PIC simulations. The zero moment of time corresponds to the moment of the peak intensity arrival at the tip of the wires.

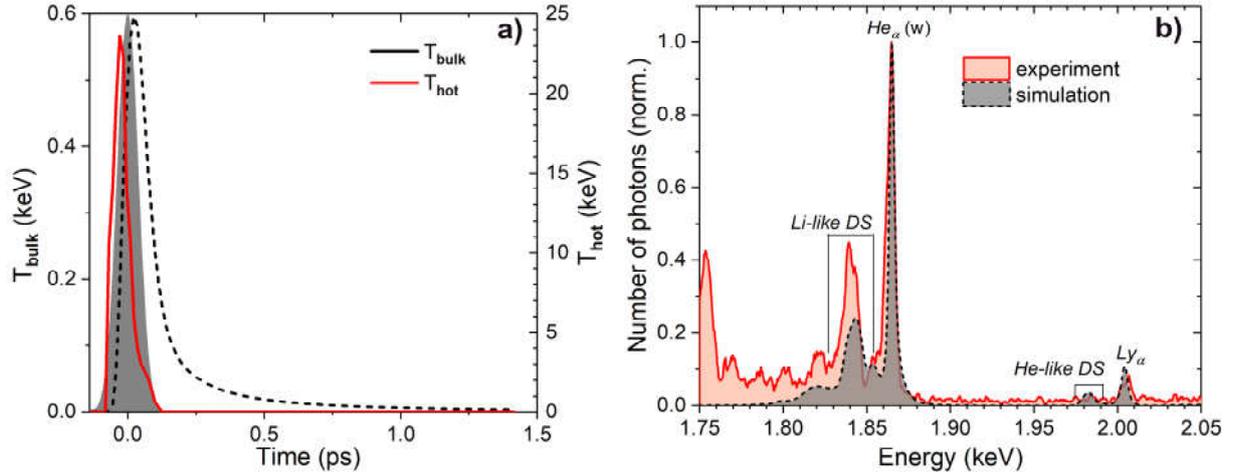

*Fig. 4* (Color on-line) a) Temporal evolution of the temperature of the "hot" (red line) and the bulk (black dashed line) electron fractions averaged over the upper 0.5 μm length of the wire. The gray shaded curve is the temporal profile of the laser intensity. b) The measured (red line) and the simulated (black dashed line) spectra (time-integrated).

In a next step we want to understand the temporal structure of the emitted X-ray radiation. FLYCHK simulations allow to get insight into the temporal evolution of the charge states and the emitted spectrum (Fig. 5a), which are governed by the evolution of the bulk electron temperature and collisional and radiative rates. Formation of He-like and H-like Si ions occurs due to the collisional ionization of $Si^{11+}$ and $Si^{12+}$, respectively. Ionization of the L-shell up to the charge state $Si^{12+}$ takes place at the rising edge of the bulk electron temperature (about 25 fs after the peak of the laser pulse intensity) (Fig. 5b). Meanwhile, ionization of the K-shell requires electrons with a kinetic energy of more than 2 keV (ionization bottleneck). Since the rates of the collisional processes in plasma drop exponentially with an increasing ratio of the binding energy $E_{bin}$ to the bulk electron temperature $T_{bulk}$, $Si^{13+}$ ions are generated with a delay in respect to the appearance of $Si^{12+}$ and approaching their highest density at roughly 75 fs (Fig. 5c). After reaching their maxima, both ion densities decrease following the $T_{bulk}$-behavior. The comparison of the maximal ion fractions obtained from time-dependent FLYCHK-simulations with a steady-state distribution at $T_{bulk}$=600 eV shows that the fraction of $Si^{13+}$ (8%, Fig. 5c) is one order of magnitude lower than obtained in the steady temperature case, while the fraction of the $Si^{12+}$ is twice higher (80% instead of 40%). Therefore, simulations suggest the transient character of the plasma under our experimental conditions.

The K-shell radiation ($He_\alpha$ and $Ly_\alpha$) arises mainly from the collisional excitation of the K-shell electrons followed by the radiative decay back into the ground state. The time scale of the build-up of the $Si^{12+}$ and $Si^{13+}$ ion densities and the time scale of the corresponding electron impact excitations are similar. Therefore, the temporal evolution of the $He_\alpha$ and $Ly_\alpha$ line emission starts almost simultaneously to the corresponding ion density. Neglecting ef-fects of opacity, the maximum of the emission is delayed by ~ 25 fs for $He_\alpha$ and ~ 50 fs for $Ly_\alpha$ with respect to the maximum of the corresponding ion density in accordance with the radiative decay times [22]. However, the $He_\alpha$ line resembles the second maximum in the emission related to the recombination channel of population from H-like ions. The time-dependent opacity strongly modifies the temporal profile of the line emission resulting in ~50 fs at FWHM pulse at the 1.865 keV $He_\alpha$ transition and ~150 fs at FWHM pulse at the 2 keV $Ly_\alpha$ transition (Fig. 5b,c). The expected shorter duration of the $He_\alpha$ emission is a consequence of higher opacity of the plasma at this emission wavelength. Overall, the ultrashort duration of the generated X-ray emission in relativistic interaction of ultra-high contrast, ultra-short laser pulses with solids is intrinsically related to the transient character of high density plasma. This transient nature of the plasma evolution has to be taken into account when estimating the time scale of radiation cooling processes and determining conditions for new regimes where the plasma cooling caused by plasma self-radiation happens faster than hydrodynamic expansion [39]. High electron density enables ultrafast collisional pumping. At the same time, high ionic density (in the corresponding charge state) leads to strong absorption of the emitted line radiation (opacity) and the re-emission of the absorbed photons is strongly suppressed due to the collisional de-excitation by free electrons. Thus, a delay between the maximum of the laser pulse and maximum of the K-shell radiation is expected as it is demonstrated in Fig.5. Therefore, we conclude that the high density and temperature of plasma, dynamically changing at sub-picosecond time scale, in combination with (dynamically changing) effects of opacity are determining the femtosecond duration of the X-ray line emission in contrast to picosecond duration estimated from assumptions of optically thin plasma under steady state conditions [39].

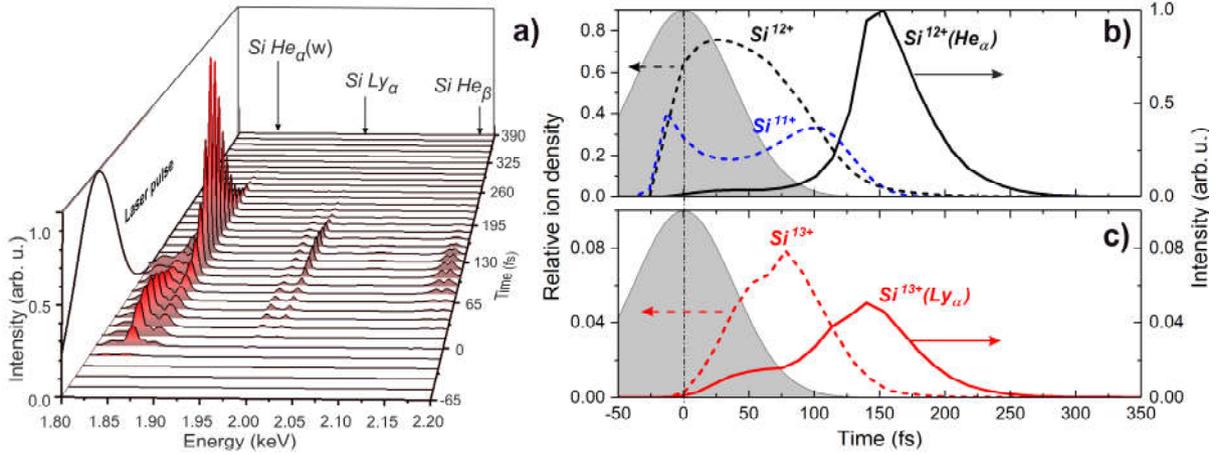

*Fig. 5* (Color on-line) a) Temporal evolution of the K-shell emission in the spectral region 1.8-2.2 keV. b) & c) Time-dependent density of $Si^{11+}$ (dashed blue line), $Si^{12+}$ (dashed black line), and $Si^{13+}$ (dashed red line) ions; intensity of $He_\alpha$ (solid black line) and $Ly_\alpha$ (solid red line) emission. The densities normalized to the density of neutral atoms. The shaded curve is the normalized laser pulse intensity.

Finally, simulations were carried out for spectra measured from a polished Si wafer to understand the differences in the emission for different morphologies of the samples. The results of spatial and temporal evolution of the bulk electron temperature and density obtained in PIC simulations are presented in Fig. 6. As follows from Fig. 6b, the plasma density for flat targets can be as high as $3 \times 10^{23}$ cm$^{-3}$ which is comparable to the density calculated for nanowires (Fig. 3b). However, the maximum bulk electron temperature is only 50 eV and reached just within a ~100 nm-thin layer near the surface. Compared to nanowires, the substantially lower electron temperature is the consequence of a strong (≈98%) reflection of the laser energy from the overcritical plasma. Simulations of the spectra by FLYCHK fail for such dense and low temperature plasma conditions. The observed spectra can be understood by extending the concept of Brunel electron heating [13], into the relativistic regime [20]. With this mechanism, electrons are efficiently accelerated near the critical density and the subsequent collision creates holes in K-shell of neutral Si atoms followed by cold $K_\alpha$ emission. Using the same long-wavelength laser system as in our experiments, an enhanced conversion efficiency for $K_\alpha$ radiation from Cu foil has been demonstrated in [19] Also we assume that the weak emission in the range of 1.747-1.827 keV (Fig. 2a), observed for all the targets and originating from intermediate charge states of Si (up to $Si^{10+}$), is generated due to inner-shell collisional ionization by energetic electrons emerging in the Brunel heating process. The difference in the yield of this emission between the nanowires and the bulk target originates from the difference in the bulk electron temperature, since mainly the bulk electrons are responsible for the evolution of the charge states in the plasma.

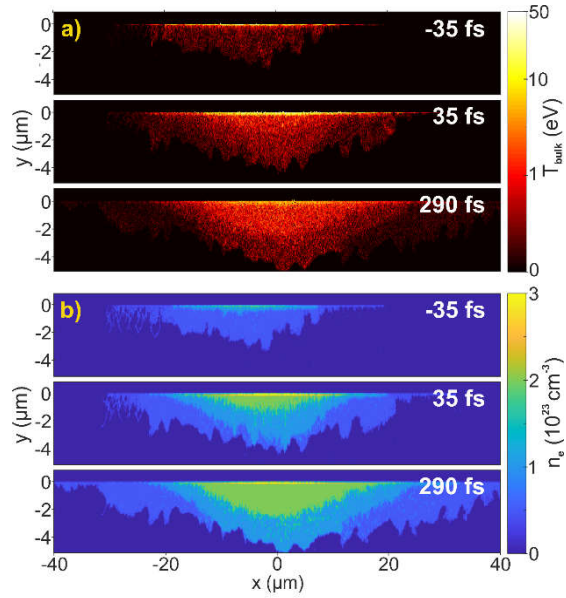

*Fig. 6* (Color on-line) The results of PIC simulations for the spatial distribution of the a) bulk electron temperature and b) electron density at the flat Si target. The zero moment of time corresponds to the moment of the peak intensity arrival at the surface boundary.

## V. Conclusions

Relativistic interaction of high temporal contrast, femtosecond laser pulses with solids is investigated experimentally and numerically in a new regime drawing on long (mid-IR) wavelength laser radiation. In this interaction regime the laser field ionization and creation of the pre-plasma on the rising edge of the laser pulse are strongly suppressed. As a consequence, the maximum of the laser pulse interacts with solid matter and ionization has predominantly collisional character. Comparing the experimentally measured X-ray emission spectra with the results of numerical simulations confirms the importance of nanowire arrays as solid target for overcoming the problem of the low critical plasma density at mid-IR wavelengths enabling very efficient absorption of the laser pulse energy (≈80%). As a result, plasma with an electron density up to $6\times10^{23}$ cm$^{-3}$ is generated, which is a record high density and is about more than 1000-times the critical density. The high electron density of the created plasmas ensures high collision rates that govern a prompt ionization of the target material up to H-like charge state ($Si^{13+}$) and 0.6 keV peak bulk electron temperature.

Using NW targets and mid-IR laser pulses we are able to generate solid density plasmas with keV-level bulk temperatures using only 20 mJ energy per laser pulse, in contrast to Joule-energy UV sources used so far to generate plasmas with similar parameters [23, 24]. Such relaxed requirements for laser energy pave the way to the experiments at high repetition rates, which is very promising for applications in laser driven nuclear physics. For instance, the cross-sections for the NEEC-reactions (Nuclear Excitation by Electron Capture) in the hot plasma will be 3-4 orders of magnitude higher compared to the "cold" XFEL-plasma (X-Ray Free-Electron Laser) case as reported in [40, 41]. Also, the simulations of the temporal evolution of the $He_\alpha$ and $Ly_\alpha$ emission suggest that ultra-high temporal contrast, relativistically intense femtosecond laser pulses enable ultrashort (≤150 fs) X-ray pulses. Such pulses are very attractive for using them for time-resolved X-ray radiography of high density plasmas [42].

**Acknowledgements**

We would like to thank N. E. Andreev for fruitful discussions. Financial support has been provided by the Federal Ministry of Education and Research, Germany grant no 05P15SJFA1. S.A., V.S., A.P., A.B. acknowledge financial support of the Austrian Science Fund (FWF), projects SFB "NextLite" F49-03 and P 27491-N27.

**References**

[1] S. Suckewer, P. Jaegle, "X-ray laser: past, present and future," Las. Phys. Lett. **6**, 411 (2009).
[2] V. N. Shlyaptsev, J. J. Rocca, M. Grisham, G. Avaria, F. Tomasel, "Prospective schemes for next generation X-ray lasers," Proc. SPIE **7451**, 745103 (2009).
[3] R. A. Smith, V. Barrow, J. Edwards, G. Kiehn, O. Willi, "High density plasmas for recombination X-ray lasers," Appl. Phys. B **50**, 187 (1990).
[4] L.J. Perkins, B.G. Logan, M.D. Rosen, M.D. Perry, T. Diaz de la Rubia, N.M. Ghoniem, T. Ditmire, P.T. Springer, S.C. Wilks, "The investigation of high intensity laser driven micro neutron sources for fusion materials research at high fluence", Nucl. Fus. **40**, 1 (2000).
[5] A. Curtis, C. Calvi, J. Tinsley, R. Hollinger, V. Kaymak, A. Pukhov, S. Wang, A. Rockwood, Y. Wang, V. N. Shlyaptsev, and J. J. Rocca, "Micro-scale fusion in dense relativistic nanowire array plasmas," Nat. Commun. **9**, 1077 (2018).
[6] S. V. Bulanov, T. Zh. Esirkepov, D. Habs, F. Pegoraro, T. Tajima, "Relativistic laser-matter interaction and relativistic laboratory astrophysics," Eur. Phys. J. D **55**, 483 (2009).
[7] S. Fujioka, H. Takabe, N. Yamamoto, D. Salzmann, F. Wang, H. Nishimura, Y. Li, Q. Dong, S. Wang, Yi Zhang, Y.-J. Rhee, Y.-W. Lee, J.-M. Han, M. Tanabe, T. Fujiwara, Y. Nakabayashi, G. Zhao, J. Zhang, K. Mima, "X-ray astronomy in the laboratory with a miniature compact object by laser driven implosion", Nat. Phys. **5**, 821 (2009).
[8] G. Mourou, Z. Chang, A. Maksimchuk, J. Nees, S. V. Bulanov, V. Yu. Bychenkov, T. Zh. Esirkepov, N. M. Naumova, F. Pegoraro, H. Ruhl, "On the design of experiments for the study of relativistic nonlinear optics in the limit of single-cycle pulse duration and single-wavelength spot size" Plasma´Phys. Rep. 28, 12 (2002).
[9] M. N. Polyanskiy, M. Babzien, I. V. Pogorelsky, "100-terawatt $CO_2$ laser: design and current status," AIP Conf. Proc. **1777**, 110006 (2016).

[10] S. Tochitsky, F. Fiuza, C. Joshi, "Prospects and directions of $CO_2$ laser-driven accelerators," AIP Conf. Proc. **1777**, 020005 (2016).
[11] G. Andriukaitis, T. Balčiūnas, S. Ališauskas, A. Pugžlys, A. Baltuška, T. Popmintchev, M.-C. Chen, M. M. Murnane, H. C. Kapteyn, "90 GW peak power few-cycle mid-infrared pulses from an optical parametric amplifier", Opt. Lett. **36**, 2755 (2011).
[12] V. Shumakova, S. Ališauskas, P. Malevich, C. Gollner, A. Baltuška, D. Kartashov, A. M. Zheltikov, A. V. Mitrofanov, A. A. Voronin, D. A. Sidorov-Biryukov, and A. Pugžlys, "Filamentation of mid-IR pulses in ambient air in the vicinity of molecular resonances", Opt. Lett. **43**, 2185-2188 (2018).
[13] F. Brunel, "Not-so-resonant, resonant absorption," Phys, Rev. Lett. 59, 52 (1987).
[14] P. Gibbon, A. R. Bell, "Collisionless absorption in sharp-edged plasmas," Phys. Rev. Lett. **68**, 1535 (1992).
[15] S. C. Wilks, W. L. Kruer, "Absorption of ultrashort, ultra-intense laser light by solids and overdence plasmas," IEEE J. Quant. Electr. **33**, 1954 (1997).
[16] P. Mulser, S.-M. Weng, T. Liseykina, "Analysis of the Brunel model and resulting hot electron spectra", Phys. Plasmas **19**, 043301 (2012).
[17] T. Liseykina, P. Mulser, M. Murakami, "Collisionless absorption, hot electron generation and energy scaling in intense laser-target interaction", Phys. Plasmas **22**, 033302 (2015).
[18] J. Wu, C. Guo, "Wavelength effects on strong-field single electron ionization," Adv. Stud. Theor. Phys. **2**, 271 (2008).
[19] J. Weisshaupt, V. Juvé, M. Holtz, ShinAn Ku, M. Woerner, T. Elsaesser, S. Ališauskas, A. Pugžlys, A. Baltuška, "High-brightness table-top hard X-ray source driven by sub-100-femtosecond mid-infrared pulses," Nat. Photonics **8**, 927–930 (2014).
[20] J. Weisshaupt, V. Juvé, M. Holtz, M. Woerner, T. Elsaesser, "Theoretical analysis of hard X-ray generation by nonperturbative interaction of ultrashort light pulses with a metal," Struc. Dyn. **2**, 024102 (2015).
[21] A. Pukhov, "Three-dimensional electromagnetic relativistic particle-in-cell code VPL (Virtual Laser Plasma Lab)," J. Plasma Phys. **61**, 425 (1999).
[22] H.-K. Chung, M.H. Chen, W.L. Morgan, Yu. Ralchenko, and R. W. Lee, "FLYCHK: Generalized population kinetics and spectral model for rapid spectroscopic analysis for all elements," High Energy Dens. Phys. **1**, 3 (2005).
[23] M. A. Purvis, V. N. Shlyaptsev, R. Hollinger, C. Bargsten, A. Pukhov, A. Prieto, Y. Wang, B. M. Luther, L. Yin, S. Wang, J. J. Rocca, "Relativistic plasma nanophotonics for ultrahigh energy density physics," Nat. Photonics **7**, 796 (2013).
[24] C. Bargsten, R. Hollinger, M. G. Capeluto, V. Kaymak, A. Pukhov, S. Wang, A. Rockwood, Y. Wang, D. Keiss, R. Tommasini, R. London, J. Park, M. Busquet, M. Klapisch, V. N. Shlyaptsev, and J. J. Rocca, "Energy penetration into arrays of aligned nanowires irradiated with relativistic intensities: scaling to terabar pressures," Science Adv. **3**, 1 (2017).
[25] G. Mourou, "The ultrahigh-peak-power laser: present and future," Appl. Phys. B., **65**, 205 (1997).
[26] S. Kroker, T. Käsebier, S. Steiner, E.-B. Kley, A. Tünnermann, "High efficiency two-dimensional grating reflectors with angularly tunable polarization efficiency," App. Phys. Lett. **102**, 161111 (2013).
[27] C. Hahn, G. Weber, R. Märtin, S. Höfer, T. Kämpfer, and Th. Stöhlker, "CdTe Timepix detectors for single-photon spectroscopy and linear polarimetry of high-flux hard x-ray radiation," Rev. Sci. Instrum. **87**, 043106 (2016).
[28] P. P. Rajeev, P. Taneja, P. Ayyub, S. S. Sandhu, and G. R. Kumar, "Metal nanoplasmas as bright source of hard X-ray pulses," Phys. Rev. Lett. **90**, 115002 (2003).

[29] T. Nishikawa, S. Suzuki, Y. Watanabe, O. Zhou, and H. Nakano, "Efficient water-window X-ray pulse generation from femtosecond-laser-produced plasma by using a carbon nanotube targets," Appl. Phys. B **78**, 885 (2004).

[30] S. Mondal, I. Chakraborty, S. Ahmad, D. Carvalho, P. Singh, A. D. Lad, V. Narayanan, P. Ayyub, G. R. Kumar, J. Zheng, and Z. M. Sheng, "Highly enhanced hard X-ray emission from oriented metal nanorod arrays excited by intense femtosecond laser pulses," Phys. Rev. B **83**, 035408 (2011).

[31] "X-ray data booklet," edited by A. C. Thompson and D. Vaughan (Lawrence Berkeley National Laboratory, University of California, 2009), 3rd Edition, 1-16.

[32] Md. R. Khan and M. Karimi, "$K_\beta/K_\alpha$ ratios in energy-dispersive X-ray emission analysis," X-ray spectrom. **9**, 1 (1980).

[33] G. H. McCall, "Calculation of x-ray bremsstrahlung and characteristic line emission produced by a Maxwellian electron distribution," J. Phys. D **15**, 823 (1982).

[34] A. Macchi, M. Borghesi, M. Passoni, "Ion acceleration by superintense laser-plasma interaction," Rev. Mod. Phys. **85**, 751 (2013).

[35] M. Roth, M. Schollmeier, "Ion acceleration—Target Normal Sheath Acceleration," edited by B. Holzer (CERN Yellow Reports, 2016), Vol. 1, p. 231.

[36] M. Sherlock, "Universal scaling of the electron distribution function in one-dimensional simulations of relativistic laser-plasma interactions," Phys. Plasmas **16**, 103101 (2009).

[37] O. N. Rosmej, Z. Samsonova, S. Höfer, D. Kartashov, C. Arda, D. Khaghani, A. Schoenlein, S. Zähter, A. Hoffmann, R. Loetzsch, I. Uschmann, M. E. Povarnitsyn, N. E. Andreev, L. P. Pugachev, M. C. Kaluza, and C. Spielmann, "Generation of keV hot near-solid density plasma states at high contrast laser-matter interaction," Phys. Plasmas **25**, 083103 (2018).

[38] A. M. Perelomov, V. S. Popov, M. V. Terent'ev, "Ionization of atoms in an alternating electric field," Sov. Phys. – JETP **23**, 924 (1966).

[39] R. Hollinger, C. Bargsten, V. N. Shlyaptsev, V. Kaymak, A. Pukhov, M. G. Capeluto, S. Wang, A. Rockwood, Y. Wang, A. Townsend, A. Prieto, P. Stockton, A. Curtis, and J. J. Rocca, "Efficient picosecond x-ray pulse generation from plasmas in the radiation dominated regime," Optica **4**, 1344-1349 (2017).

[40] J. Gunst, Y. A. Litvinov, C. H. Keitel, and A. Pálffy, "Dominant secondary nuclear photoexcitation with the X-Ray free-electron laser," Phys. Rev. Lett. **112**, 082501 (2014).

[41] J. Gunst, Y. Wu, N. Kumar, C. H. Keitel, and A. Pálffy, "Direct and secondary nuclear excitation with X-ray free-electron lasers," Phys. Plasmas **22**, 112706 (2015).

[42] A. Morace, L. Fedeli, D. Batani, S. Baton, F. N. Beg, S. Hulin, L. C. Jarrot, A. Margarit, M. Nakai, P. Nakatsutsumi, P. Nicolai, N. Piovella, M. S. Wei, X. Vaisseau, L. Volpe, and J. J. Santos, "Development of x-ray radiography for high energy density physics," Phys. Plasmas **21**, 102712 (2014).